\documentclass[journal]{IEEEtran}
\usepackage{amssymb, amsmath, amsthm, graphicx, enumerate}
\usepackage{multirow}
\usepackage{setspace}
\usepackage{algorithm}
\usepackage{algpseudocode}
\usepackage{array}
\usepackage{subfigure}
\usepackage{float}
\begin{document}
\newcommand{\fwidthtwo}{0.3}
\newcommand{\fwidththree}{0.2}
\newcommand{\fwidthgraph}{0.4}

\makeatletter
\newcommand\figcaption{\def\@captype{figure}\caption}
\newcommand\tabcaption{\def\@captype{table}\caption}
\makeatother

\title{\baselineskip=0.6\baselineskip Learning Decorrelated Hashing Codes for Multimodal Retrieval}
\author{Dayong~Tian}
\IEEEtitleabstractindextext{
\begin{abstract}
In social networks, heterogeneous multimedia data correlate to each other, such as videos and their corresponding tags in YouTube and image-text pairs in Facebook. Nearest neighbor retrieval across multiple modalities on large data sets becomes a hot yet challenging problem. Hashing is expected to be an efficient solution, since it represents data as binary codes. As the bit-wise XOR operations can be fast handled, the retrieval time is greatly reduced. Few existing multimodal hashing methods consider the correlation among hashing bits. The correlation has negative impact on hashing codes. When the hashing code length becomes longer, the retrieval performance improvement becomes slower. In this paper, we propose a minimum correlation regularization (MCR) for multimodal hashing. First, the sigmoid function is used to embed the data matrices. Then, the MCR is applied on the output of sigmoid function. As the output of sigmoid function approximates a binary code matrix, the proposed MCR can efficiently decorrelate the hashing codes. Experiments show the superiority of the proposed method becomes greater as the code length increases.
\end{abstract}
}
\maketitle
\IEEEdisplaynontitleabstractindextext
\IEEEpeerreviewmaketitle

\section{Introduction}
Multimodal hashing which embeds data to binary codes is an efficient tool for retrieving heterogeneous but correlated multimedia data, such as image-text pairs in Facebook and video-tag pairs in Youtube. Unlike real vectors used in traditional retrieval methods~\cite{real1}\cite{real2}\cite{real3}\cite{real4}, binary codes can greatly reduce the storage requirement and computation costs of nearest neighbors search. \\
\indent Existing multimodal hashing methods can be classified into supervised and unsupervised ones according to whether the label information is used. Unsupervised multimodal hashing aims at preserving the Euclidean distances between each pair of data. Inter-media hashing (IMH) seeks a common Hamming space in which binary codes preserve inter-media consistency and intra-media consistency~\cite{mIMH}. To avoid the large-scale graph which needs to compute and store the pairwise distances, linear cross-modal hashing (LCMH)~\cite{LCMH} computes distances between each training data point and a small number of cluster centers. Collective matrix factorization hashing (CMFH)~\cite{CMFH} uses collective matrix factorization on each modality to learn a unified hashing codes. Zhu~\emph{et al.}~\cite{mobile} incorporates texts to facilitate unsupervised image retrieval. Although multimodal data is used in~\cite{mobile}, it is designed for unimodal retrieval task.\\
\indent By incorporating label information, supervised hashing can preserve semantic information and achieve higher accuracy. Cross-modality similarity-sensitive hashing (CMSSH)~\cite{CMSSH} treats hashing as a binary classification problem. Cross-view hashing (CVH)~\cite{CVH} assumes the hashing codes be a linear embedding of the original data points. It substitutes the code matrix by this embedding. The objective function is a weighted summation of that of spectral hashing (SH)~\cite{SH} on each modality. Multilatent binary embedding (MLBE)~\cite{MLBE} treats hashing codes as the binary latent factors in the proposed probabilistic model and maps data points from multiple modalities to a common Hamming space. Semantics-preserving hashing (SePH)~\cite{SePH} learns the hashing codes by minimizing the KL-divergence of the probability distribution in Hamming space from that in semantic space. CMSSH, MLBE and SePH need to compute the affinities of all data points, which makes it intractable for large data set. Semantic correlation maximization (SCM)~\cite{SCM} circumvents this by learning only one bit each time and the explicit computation of affinity matrix is avoided through several mathematical manipulations. Multimodal discriminative binary embedding (MDBE) models~\cite{MDBE} hashing as a minimization problem. There are two main terms in its formulation. One term indicates different modalities and the labels can be embedded to the same latent space, while the other one indicates the embedded modalities can be further embedded as the labels. $l2$-norm is used to regularize the linear embedding matrix. SCM and MDBE discard the uncorrelation property of the code matrix or embedding matrix, which makes their performance improve slowly as code length increases.\\
\indent Wang~\emph{et al.}\cite{aaaiorth} introduces an orthogonality regularization (OR) to their deep neural network (DNN) hashing model. They use Restricted Boltzmann Machine (RBM) for image and text data. Each layer of RBM can be represented as a nonlinear activation function of a linear transformation of the input. The OR is applied on the weight matrix of each layer. Wang~\emph{et al.} argue that the proposed OR can lead to an orthogonal code matrix when data matrices are orthogonal. This assumption is unreasonable in real application. In this paper, we will briefly analyze the properties of this OR and demonstrate that it is only suitable for some linear hashing models. Deep cross-modal hashing (DCMH)~\cite{DCMH} employs different types of DNN for different modalities. For example, convolutional neural network (CNN) is used for images while full connected neural network is used for text. The orthogonality of hashing codes is neglected.\\
\indent In this paper, we propose a hashing method named decorrelated multimodal hashing (DMH). First, a sigmoid function is applied on the linear transformations of original data points to map different modalities into a common code matrix. Then, we devise a minimum correlation regularization (MCR) to improve the retrieval performance on long-bit experiments. Unlike aforementioned orthogonality constraints or regularizations that are usually applied on the linear transformation matrices, the proposed MCR is applied on the sigmoid function. Because the output of sigmoid function approximates a binary code and the hashing code matrix directly depends on the quantization of it, the propose MCR works better on decorrelating hashing codes. \\
\indent We do not use the term ``orthogonality'' because the maximum number of mutual orthogonal vectors is equal to the dimension of them and an orthogonal linear transformation does not exist when the rank of a data matrix is less than that of its code matrix. For instance, if an $N\times d$ data matrix is encoded as an $N\times c$ code matrix where $N$ is the number of data and $d<c$, the dimension of the linear transformation matrix $W$ should be $d\times c$. Because we cannot find $c$ $d$-dimensional column vectors, an orthogonal $W$ does not exist. In Subsection~\ref{subsec:ortho}, we will prove that when $d+1<c$, the output matrix of sigmoid function cannot be orthogonal and hence the orthogonality of code matrix cannot be even approximated.\\
\indent The rest of this paper is organized as follows. The related works are reviewed in Section~\ref{sec:related}. In Section~\ref{sec:premodel}, we, step by step, derive our model from a widely used unimodal hashing method, iterative quantization (ITQ)~\cite{ITQ}. The discussions on parameter settings and optimization algorithms are also given in Section~\ref{sec:premodel}. Experimental results are reported in Section~\ref{sec:experiment}. We conclude this paper in Section~\ref{sec:conclusion}
\section{Related Works}
\label{sec:related}
Some well-known multimodal hashing models are related to some classical unimodal ones. Hence, in this section, unimodal hashing models will be firstly reviewed and then we will discuss some representative multimodal hashing models and their relations to unimodal ones.\\
\subsection{Unimodal Hashing}
Unimodal hashing can be divided into two categories according to their dependence on data. Locality-sensitive hashing (LSH)~\cite{LSH} and its kernelized version~\cite{KLSH}\cite{CLSH} are well-known data-independent unsupervised unimodal hashing methods. Due to randomized hashing, LSH demands more bits per hashing table~\cite{IMH}.\\
\indent Spectral hashing (SH)~\cite{SH}, one of the most popular and pioneering data-dependent unimodal hashing methods, generate hashing codes by solving a relaxed mathematical problem to avoid computing the affinity matrix that requires calculating and storing pairwise distances of the whole data set~\cite{GHS}. The authors argued that two constraints for a good code matrix are orthogonality and balance, either of which leads to an NP-hard problem. In the following works, balance is generally neglected and orthogonality constraint is relaxed or neglected, too.\\
\indent Anchor graph hashing (AGH)~\cite{AGH} substitutes the affinity matrix in SH by constructing the a highly sparse one using several anchor points. Discrete graph hashing (DGH)~\cite{DGH} incorporates a relaxed orthogonality constraint into AGH to improve the performance on long-bit experiments.\\
\indent Methods based on linear transformations, such as principal component analysis (PCA)~\cite{PCA}, attract wide interests due to their effectiveness and computation efficiency. ITQ rotates the projection matrix obtained by PCA to minimize the quantization loss. Isotropic hashing (IsoH)~\cite{HH}, harmonious hashing (HH)~\cite{HH} and ok-means~\cite{okmeans} are derived from ITQ. IsoH equalizes the importance of principal components. HH puts an orthogonal constraint on an auxiliary variable for the code matrix. ok-means rotates the data matrix to minimize the quantization loss. ITQ, IsoH and HH depends on principal components whose maximum number is no larger than the minimum dimension of data matrix. Hence, they cannot generate hashing codes longer than the data dimension. Despite of PCA, other linear transformations can be used, such as Linear Discriminant Analysis (LDA)~\cite{LDAHashing}. Unlike these pre-computed transformation matrix, neighborhood discriminant hashing~\cite{NDH} calculates the transformation matrix during the iterative minimization procedure.\\
\indent Inductive manifold hashing~\cite{IMH} embeds some special samples into lower dimensional space and the embeddings of remaining samples are calculated by a linear combination of those special samples. The coefficients of the linear combination are the probabilities that a sample belongs to those special samples.\\
\indent All aforementioned unimodal hashing models cannot generate balanced code matrix. Spherical hashing (SpH)~\cite{SpH} and global hashing system (GHS)~\cite{GHS} quantize the distance between a data point and a special point. The closer half to a special point is denoted as 1 while the further half is denoted as 0. Therefore, a balanced matrix can be easily generated. Their major difference is on how to find these special points. SpH uses a heuristic algorithm while GHS treats it as a satellite distribution problem of the Global Positioning System (GPS).\\
\subsection{Multimodal Hashing}
Multimodal hashing models can be classified into unsupervised and supervised ones. Unsupervised multimodal hashing tries to preserve the Euclidean data structure by binary codes. Inter-media hashing~\cite{IMH} learns hashing function by linear regression. IMH models intra-media consistency in a similar way of SH. Like what AGH has done to SH, linear cross-media hashing (LCMH)~\cite{LCMH} uses the distances between each data point and each cluster centroid to construct a sparse affinity matrix. Collective matrix factorization hashing (CMFH)~\cite{CMFH} can be treated as an extension of NDH. For each modality, CMFH consists of two terms: (1) calculating a transformation matrix for the data matrix to match the code matrix through minimizing quantization loss, and (2) calculating a transformation matrix for the code matrix to match the data matrix through minimizing squared error. Latent semantic sparse hashing~\cite{LSSH} is an extension of CMFH and its basic idea is similar to HH that imposes the orthogonality constraint on an auxiliary variable. LSSH imposes the sparse regularization on an auxiliary variable in the latent space. Shen~\emph{et al.}~\cite{semipaired} proposed a cross-view hashing method for semi-paired data. It jointly learns a correlated representation for each modality and hashing functions. It rotates the hashing code matrix to match the correlated representation matrices. Hence, it can be seen as an extension of ok-means.\\
\indent By incorporating label information, supervised multimodal hashing can achieve higher accuracy than unsupervised ones. Cross-view Hashing (CVH) extends SH by minimizing the weighted average Hamming distances of hashing codes of training data pairs. The minimization is solved as a generalized eigenvalue problem. However, the performance of CVH decreases with increasing bit number, because most of the variance is contained in the top few eigenvectors~\cite{MDBE}. Multimodal discriminative binary embedding (MDBE) models~\cite{MDBE} derives from CMFH. It is comprised of (1) transforming data matrix to label matrix and (2) transforming label matrix and data matrix to a latent space.
\section{Methodology}
\label{sec:premodel}
Terms ``view'' and ``modality'' are discriminated in some literatures~\cite{MDBE}. Multiple views of data refers to different type of features of one modality, e.g. SIFT~\cite{SIFT} and GIST~\cite{GIST} features for images. However, we use these two words interchangeably since our method can be used in either situations as long as the data are represented by real matrices. \\
\indent First, Let us define the used notations. Suppose that $\mathbf{X}^i$ is the $i$-th view matrix of the data and $\mathbf{X}^i=\left[\mathbf{x}^i_1,\ldots,\mathbf{x}^i_n\right]^\top$, where $\mathbf{x}^i_m\in\mathbb{R}^{d_i}$, $n$ is the number of data points and $i=1,\ldots,g$. A binary code corresponding to the $m$-th data is defined by a row vector $b_m=\{0,1\}^c$, where $c$ is the code length and the code matrix $\mathbf{B}=\left[\mathbf{b}_1^\top,\ldots,\mathbf{b}_n^\top\right]^\top$. $h^i\left(\mathbf{X}^i\right.)$, the hashing function for the $i$-th view matrix, embeds $\mathbf{X}^i$ into a binary code matrix.\\
\subsection{Problem Formulation}
ITQ is a successful hashing method for single view data. The formulation of ITQ is
\begin{equation}
\mathop {\arg \min }\limits_{{\bf{B}},{\bf{R}}} E=\left\| {{\bf{B}} - {\bf{XWR}}} \right\|_F^2,
\end{equation}
where $\mathbf{X}\in\mathbb{R}^{n\times d}$ is the data matrix, $\mathbf{W}\in\mathbb{R}^{d\times c}$ is obtained by principal component analysis (PCA) and $\mathbf{R}\in\mathbb{R}^{c\times c}$ is an orthogonal matrix. An intuitive multi-view extension of ITQ can be
\begin{equation}\label{eq:multiITQ}
\mathop {\arg \min }\limits_{{\bf{B}},{\bf{R}}^i} E=\sum_i{\alpha_i\left\| {{\mathbf{B}} - {\mathbf{X}^i\mathbf{W}^i\mathbf{R}^i}} \right\|_F^2},
\end{equation}
where $\alpha_i$ is a positive real constant. As the maximum number of principal components pre-computed by PCA on the $i$th view matrix is $d_i$, Eq.~\eqref{eq:multiITQ} cannot be used when $c>d_i$. We remove $\mathbf{R}^i$ from Eq.~\eqref{eq:multiITQ}. Then, we simultaneously calculate $\mathbf{W}^i\in\mathbb{R}^{d_i\times c}$ and $\mathbf{B}$ during the optimization process. This method can be modeled as
\begin{equation}\label{eq:preModel}
\mathop {\arg \min }\limits_{{\bf{B}},{\bf{W}}^i} E=\sum_i{\alpha_i\left\| {{\mathbf{B}} - {\mathbf{X}^i\mathbf{W}^i}} \right\|_F^2}.
\end{equation}
Because $\mathbf{B}$ is a binary matrix, $h^i(\mathbf{X}^i\mathbf{W}^i)=1/(1+\exp(-(\beta_i*\mathbf{X}^i\mathbf{W}^i+\mathbf{1}\mathbf{v}^i)))$ is applied to transform the values of $\beta_i*\mathbf{X}^i\mathbf{W}^i+\mathbf{1}\mathbf{v}^i$ into interval $(0,1)$, where $\mathbf{1}$ is a $n$-dimensional column vector whose elements are equal to 1. $\beta_i$ is a constant and $\mathbf{v}^i$ is a bias vector. Hence, Eq.~\eqref{eq:preModel} can be modified as following.
\begin{equation}\label{eq:preModel2}
\mathop {\arg \min }\limits_{{\bf{B}},{\bf{W}}^i,{\mathbf{v}^i}} E=\sum_i{\alpha_i\left\| {{\mathbf{B}} - \frac{1}{1+\exp\left(-(\beta_i\mathbf{X}^i\mathbf{W}^i+\mathbf{1}\mathbf{v}^i)\right)}} \right\|_F^2}.
\end{equation}
\subsection{Minimum Correlation Regularization}
\label{subsec:ortho}
The orthogonality condition for good codes~\cite{SH} is approximated by an orthogonal $\mathbf{W}$ in ITQ. However, when $c>d_i$, an orthogonal $\mathbf{W}^i$ does not exist. In this case, Wang~\emph{et al.}~\cite{aaaiorth} introduces the following regularization to decorrelate code matrix:
\begin{equation}\label{eq:mcrtmp}
\textrm{R}=\left\|{\mathbf{W}^i}^\top\mathbf{W}^i-\mathbf{I}\right\|_F
\end{equation}
\indent First, let us discuss some interesting properties of Eq.~\eqref{eq:mcrtmp}.\\
\textbf{Proposition 1}. When $c\leq d_i$, the $\mathbf{W}^i$ that minimizes Eq.~\eqref{eq:mcrtmp} is an orthogonal matrix.\\
\indent It is easy to prove \textbf{Proposition 1} by the definition of orthogonal matrix.\\
\textbf{Proposition 2}. Let the $\mathbf{W}^i$ that minimizes Eq.~\eqref{eq:mcrtmp} consists of column vectors $\mathbf{w}^i_p$ where $p=1,\ldots,c$. The angle between any pair of column vectors is equal to each other.
\begin{proof}
Let $\mathbf{V} = {\mathbf{W}^i}^\top\mathbf{W}^i$ and let $V_{pq}$ be the element in the $p$th row and $q$th column of $\mathbf{V}$. $V_{pq}$ is the inner product of $\mathbf{w}^i_p$ and $\mathbf{w}^i_q$. When $\|\mathbf{w}_p^i\|_F^2=1$, the diagonal elements of $R$ will be 0 and the angle between $\mathbf{w}^i_p$ and $\mathbf{w}^i_q$ will be $\arccos({{\mathbf{w}^i_p}^\top\mathbf{w}^i_q})$. Eq.~\eqref{eq:mcr} can be written as:
\begin{equation}
\textrm{R}=\sum_{p,q}{{\mathbf{w}^i_p}^\top\mathbf{w}^i_q},\quad p\neq q
\end{equation}
According to the inequality of arithmetic and geometric means, it can be deduced that
\begin{equation}
\frac{\sum_{p,q}{{\mathbf{w}^i_p}^\top\mathbf{w}^i_q}}{c^2-c} \geq \prod_{p,q}{\sqrt[\leftroot{-2}\uproot{2}{c^2-c}]{{\mathbf{w}^i_p}^\top\mathbf{w}^i_q}}
\end{equation}
The equality holds if and only if all ${\mathbf{w}^i_p}^\top\mathbf{w}^i_q$ are equal. That is, the angle between any pair of column vectors is equal when $\mathbf{W}^i$ minimizes Eq.~\eqref{eq:mcrtmp}.
\end{proof}
\noindent\textbf{Proposition 3}. If $\mathbf{W}^i$ minimizes Eq.~\eqref{eq:mcrtmp}, the affine transformation of $\mathbf{W}^i$, i.e. $\mathbf{W}^i\mathbf{R}$ also minimizes Eq.~\eqref{eq:mcrtmp} where $\mathbf{R}$ is an orthogonal matrix.\\
\begin{proof}
As $\mathbf{R}$ is orthogonal, we have
\begin{equation}\label{eq:8}
\left\|{\mathbf{W}^i}^\top\mathbf{W}^i-\mathbf{I}\right\|_F=\left\|\mathbf{R}^\top\left({\mathbf{W}^i}^\top\mathbf{W}^i-\mathbf{I}\right)\mathbf{R}\right\|_F
\end{equation}
Eq.~\eqref{eq:8} can be rewritten as
\begin{equation}
\left\|{\mathbf{W}^i}^\top\mathbf{W}^i-\mathbf{I}\right\|_F=\left\|\mathbf{R}^\top{\mathbf{W}^i}^\top\mathbf{W}^i\mathbf{R}-\mathbf{I}\right\|_F
\end{equation}
Here, $\mathbf{R}^\top\mathbf{R}=I$ is used in the deduction. Hence, $\mathbf{W}^i\mathbf{R}$ also minimizes Eq.~\eqref{eq:mcrtmp}.
\end{proof}
\begin{figure}[h]
\centering
\includegraphics[width = 1\linewidth]{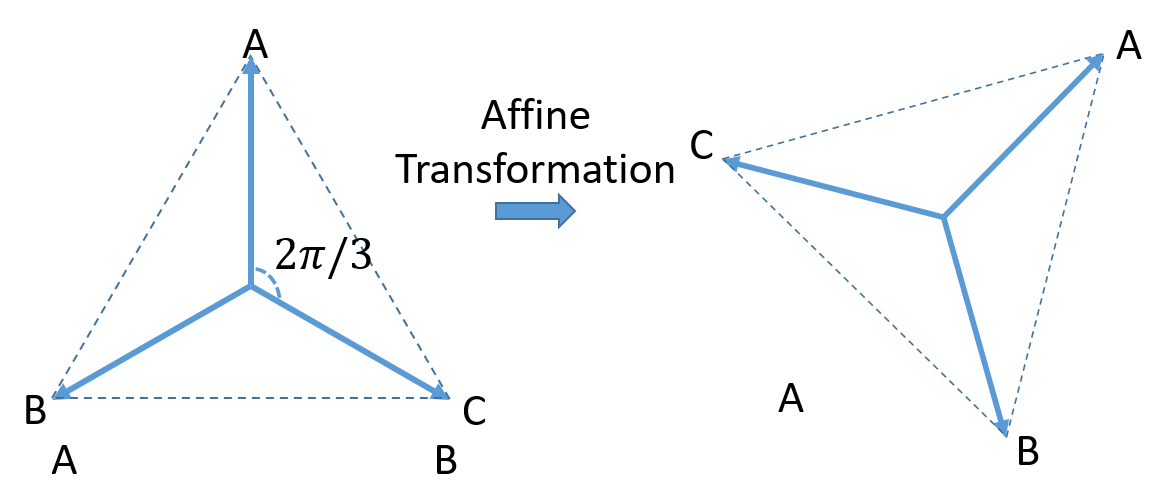}
\caption{Illustration of \textbf{Proposition 2} and \textbf{Proposition 3}. If $\mathbf{W}^i\in \mathbb{R}^{2\times 3}$, its column vectors will align with the centerlines of an equilateral triangle. The affine transformation will change the relative positions among vectors but the overall structure is kept. In the equilateral triangle, point B is transformed to the clockwise direction of point A.}
\label{fig:proposition}
\end{figure}
\indent In Fig.~\ref{fig:proposition}, we illustrate \textbf{Proposition 2} and \textbf{Proposition 3} in 2-dimensional case. Following the flowchart of ITQ, one can find $c$ $d$-dimensional vectors distributed like those in Fig.~\ref{fig:proposition} and then transform them by $\mathbf{R}$ to minimize Eq.~\eqref{eq:multiITQ}. However, the complexity of theoretically finding such vectors increases dramatically in high dimensional spaces. Wang~\emph{et al.}~\cite{aaaiorth}  use Eq.~\eqref{eq:mcrtmp} as a regularization and argue that Eq.~\eqref{eq:mcrtmp} will lead to an orthogonal code matrix when the data matrices are orthogonal. It is easy to find an example demonstrating Eq.~\eqref{eq:mcrtmp} can only be used in some linear models. For simplicity, let us consider the following model,
\begin{equation}\label{eq:counterexample}
\mathop {\arg \min }\limits_{{\bf{B}},{\bf{W}}} E=\left\|\mathbf{B} - f(\mathbf{XW})\right\|,
\end{equation}
where $\mathbf{X}$ is an orthogonal data matrix and $f(\cdot)$ is a linear or nonlinear function. Please note Eq.~\eqref{eq:counterexample} is not a unimodal hashing model, because the binary constraint is not imposed to $\mathbf{B}$. Let us suppose the dimensions of $\mathbf{B}$ and $\mathbf{X}$ are equal. According to \textbf{Proposition 1}, Eq.~\eqref{eq:mcrtmp} will lead to an orthogonal $\mathbf{W}$. If $f(\mathbf{XW})=\mathbf{XW}$, then $\mathbf{B}=\mathbf{XW}$ is also an orthogonal matrix. However, if $f(\cdot)$ is a sign function which is nonlinear, we can get a binary code matrix $\mathbf{B}=sign(\mathbf{XW})$ and Eq.~\eqref{eq:counterexample} becomes a nonlinear unimodal hashing model.  Obviously, an orthogonal $\mathbf{W}$ cannot ensure an orthogonal $\mathbf{B}$.  \\
\indent Inspired by this example, we propose the following regularization
\begin{equation}\label{eq:mcr}
\left\|\frac{f^\top(\mathbf{X},\Theta) f(\mathbf{X},\Theta)}{n}-\mathbf{I}\right\|_F,
\end{equation}
where $f(\mathbf{X},\Theta)$ is the nonlinear embedding function and $\Theta$ is the parameter set of $f$. In our proposed hashing model, i.e., Eq.~\eqref{eq:preModel2},
\begin{equation}
f(\mathbf{X}^i,\mathbf{W}^i, \mathbf{v}^i,\beta_i)=  \frac{1}{1+\exp\left(-(\beta_i\mathbf{X}^i\mathbf{W}^i+\mathbf{1}\mathbf{v}^i)\right)}.
\end{equation}
\textbf{Proposition 4}. Minimizing Eq.~\eqref{eq:mcr} cannot lead to an orthogonal $f(\mathbf{X}^i,\Theta)$ when $d_i+1<c$.
\begin{proof}
According to the definitions, we have $rank(\mathbf{X}^i)\leq d_i$, $rank(\mathbf{W}^i)\leq d_i$ and $rank(\mathbf{1}\mathbf{v}^i)\leq 1$. Hence, $rank(\mathbf{X}^i\mathbf{W}^i)\leq d_i$ and $rank(\mathbf{X}^i\mathbf{W}^i+\mathbf{1}\mathbf{v}^i)\leq d_i+1$.\\
According to \textbf{Theorem 4.2} in~\cite{nonlinearmatrix}, $rank(f(\mathbf{X}^i\mathbf{W}^i+\mathbf{1}\mathbf{v}^i))\leq d_i+1$, and hence
\begin{equation}
rank(f^\top (\mathbf{X}^i\mathbf{W}^i+\mathbf{1}\mathbf{v}^i)f (\mathbf{X}^i\mathbf{W}^i+\mathbf{1}\mathbf{v}^i))\leq d_i+1.
\end{equation}
Because $d_i+1<c$, $f^\top (\mathbf{X}^i\mathbf{W^i}+\mathbf{1}\mathbf{v}^i)f (\mathbf{X}^i\mathbf{W^i}+\mathbf{1}\mathbf{v}^i)$ cannot be equal to $\mathbf{I}$ in any cases. Hence, Minimizing Eq.~\eqref{eq:mcr} cannot lead to an orthogonal $f$.
\end{proof}
From \textbf{Proposition 4}, we can see that an orthogonal $f$ cannot be acquired when $d_i+1<c$. In this case, $f$ cannot even approximate an orthogonal matrix. Minimizing $f$ will only minimize the correlation among the column vectors of $f$. Fig.~\ref{fig:proposition} illustrates this situation.\\
\indent It is inessential to name Eq.~\eqref{eq:mcr} as ``minimum correlation regularization'' (MCR) or ``maximum uncorrelation regularization''. Since Eq.~\eqref{eq:mcr} will be added into our hashing model which is formulated as a minimization problem,  we use the former one to keep literal consistency.
\subsection{Decorrelated Multimodal Hashing}
\indent In our implementation, we found that subtracting identity matrix is somewhat redundant, so MCR can be simplified as:
\begin{equation}
\left\|\frac{f^\top(\mathbf{X},\Theta) f(\mathbf{X},\Theta)}{n}\right\|_F.
\end{equation}
It is unnecessary to worry about the diagonal elements of $f^\top f$ will be zeros during the proposed minimization procedure, because as long as all variables are randomly initialized, it is nearly impossible for gradient descent algorithm to reach a solution that all variables are zero. \\
\indent Adding MCR to Eq.~\eqref{eq:preModel2} leads to the following model.
\begin{equation}\label{eq:model}
\mathop {\arg \min }\limits_{{\bf{B}},{\bf{W}}^i,{\mathbf{v}^i}} E=\sum_i{\alpha_i\left(\left\| \mathbf{B} - \frac{1}{1+\mathbf{A}^i} \right\|_F^2+\gamma_i\left\|\left(\frac{1}{1+\mathbf{A}^i}\right)^\top\frac{1}{1+\mathbf{A}^i}\right\|_F\right)},
\end{equation}
where $\gamma_i$ is a positive real constant, and
\begin{equation}
\mathbf{A}^i = \exp\left(-(\beta_i\mathbf{X}^i\mathbf{W}^i+\mathbf{1}\mathbf{v}^i)\right).
\end{equation}
\subsection{Optimization}
Eq.~\eqref{eq:model} is minimized by iterative minimization. Take the partial derivative with respect to $\mathbf{B}$, resulting in
\begin{equation}\label{eq:updateB}
\frac{\partial E}{\partial \mathbf{B}}=2\sum_i{\alpha_i\mathbf{B}}-2\sum_i{\frac{\alpha_i}{1+\exp\left(-(\beta_i\mathbf{X}^i\mathbf{W}^i+\mathbf{1}\mathbf{v}^i)\right)}}
\end{equation}
Setting Eq.~\eqref{eq:updateB} as 0, we can derive that
\begin{equation}
\mathbf{B}=\frac{1}{\sum_i{\alpha_i}}\sum_i{\frac{\alpha_i}{1+\exp\left(-(\beta_i\mathbf{X}^i\mathbf{W}^i+\mathbf{1}\mathbf{v}^i)\right)}}
\end{equation}
$\mathbf{B}$ is rounded in each iteration to ensure $\mathbf{B}\in\{0,1\}^{n\times c}$.\\R
Take the partial derivative with respect to $\mathbf{v}^i$, resulting in
\begin{equation}\label{eq:updatab}
\frac{\partial E}{\partial \mathbf{v}^i}=\frac{2\alpha_i}{n}\left(\mathbf{C}^i-\mathbf{B}+\gamma_i\left(\mathbf{C}^i\right)^\top\mathbf{C}^i\right)\circ \left(\mathbf{A}^i\circ\frac{1}{(1+\mathbf{A}^i)^2}\right)
\end{equation}
\begin{equation}
\mathbf{C}^i=\frac{1}{1+\mathbf{A}^i}
\end{equation}
In Eq.~\eqref{eq:updatab}, ``$\circ$'' means element-wise multiplication. The division and square are also element-wise. The partial derivative with respect to $\mathbf{W}^i$ is
\begin{equation}\label{eq:updateW}
\frac{\partial E}{\partial \mathbf{W}^i}=2\alpha_i\beta_i{\mathbf{X}^i}^\top\left(\mathbf{C}^i-\mathbf{B}+\frac{\gamma_i}{n}\mathbf{C}^i\left(\mathbf{C}^i\right)^\top\mathbf{C}^i\right)\circ \left(\mathbf{A}^i\circ\frac{1}{(1+\mathbf{A}^i)^2}\right)
\end{equation}
The prototype of the proposed training method is shown in \textbf{Algorithm~\ref{alg:1}}. In Subsection~\ref{subsec:id}, the parameter settings and details for efficient implementation are discussed.
\begin{algorithm}
\caption{~~The Prototype of the Proposed Training Method}
\label{alg:1}
\begin{algorithmic}[1]
\Require $\alpha_i$, $\beta_i$, $\triangle t$, $\mathbf{X}^i$
\While {$E$ not converged}
\State Update $\mathbf{B}$ using Eq.~\eqref{eq:updateB}.
\State $\mathbf{v}^i\leftarrow\mathbf{v}^i-\triangle t\cdot{\partial E}/{\partial \mathbf{v}^i}$
\State $\mathbf{W}^i\leftarrow\mathbf{W}^i-\triangle t\cdot{\partial E}/{\partial \mathbf{W}^i}$
\EndWhile
\Ensure $\bf{B},\mathbf{W}^i,\mathbf{v}^i$
\end{algorithmic}
\end{algorithm}
\subsection{Implementation details}
\label{subsec:id}
$\alpha_i$ is weight for $i$th view. We set $\alpha_i$ as 10 for label view and 1 for any other views. $\beta_i$ is used to re-scale the view matrix. We empirically found that the proposed method achieves the best performance when the values of the re-scaled view matrix are in the interval $\left[0,255\right]$. For instance, in the NUS-WIDE dataset~\cite{NUS-WIDE}, images are represented by 500-dimensional bag-of-visual-words SIFT feature vectors whose values are in $\left[0,255\right]$, texts are represented by 1000-dimensional index vectors whose values are 0 or 1 and labels are 10-dimensional index vectors. Hence, we set $\beta$ as 1, 255 and 255 for image view matrix, text view matrix and label view matrix, respectively. To improve computation efficiency, $\beta_i$ is multiplied with $\mathbf{X}_i$ before the iteration starts.\\
\indent We set the maximum iteration times as $K$. $\triangle t$ linearly decreases from $k_s$ to $k_e$ by $K$ iterations, i.e., in the $k$-th iteration, $\triangle t = k_s - (k_s-k_e)k/K$. \\
\indent For large dataset, the first term in Eq.~\eqref{eq:model} is too large, which makes $\gamma_i$ and $\triangle t$ difficult to be determined. We normalize the gradients so that we can fix $\gamma_i$ and $\triangle t$ settings for all our experiments. The efficient version of the proposed method is given in \textbf{Algorithm}~\ref{alg:2}
\begin{algorithm}
\caption{~~The Proposed Training Method}
\label{alg:2}
\begin{algorithmic}[1]
\Require $\alpha_i$, $\beta_i$, $\triangle t$, $\mathbf{X}^i$, $k$, $k_s$, $k_e$, $K$
\While {$E$ not converged and $k < K$}
\State $\triangle t = k_s - (k_s-k_e)k/K$
\State Update $\mathbf{B}$ using Eq.~\eqref{eq:updateB}.
\State $\mathbf{v}^i\leftarrow\mathbf{v}^i-\triangle t\cdot{\partial E}/{\partial \mathbf{v}^i}$
\State $\mathbf{W}^i\leftarrow\mathbf{W}^i-\triangle t\cdot\frac{{\partial E}/{\partial \mathbf{W}^i}}{\|{\partial E}/{\partial \mathbf{W}^i}\|}$
\State $k\leftarrow k+1$
\EndWhile
\Ensure $\bf{B},\mathbf{W}^i,\mathbf{v}^i$
\end{algorithmic}
\end{algorithm}
\begin{table*}[!h]
\caption{MAP results on MIRFlickr and NUS-WIDE data sets.}
\label{tb:map}
\begin{center}
\begin{tabular}{|c|c|c	c	c	c	c|}
\hline
\multirow{ 2}{*}{Task} & \multirow{2}{*}{Method} & \multicolumn{5}{|c|}{\bf MIRFlickr} \\ \cline{3-7} & & 16 bits & 32 bits & 64 bits & 96 bits & 128 bits\\ \hline
\multirow{6}{*}{Image-Text}& CMSSH & 0.5966 & 0.5674 & 0.5581 & 0.5692 & 0.5701 \\ 
&CVH &0.6591 & 0.6145 & 0.6133 & 0.6091 & 0.6052 \\ 
&SCM &0.6251 &0.6361 &0.6417 &0.6446 &0.6480\\ 
&SePH &0.6505 &0.6447 &0.6453 &0.6497 &0.6612\\ 
&MDBE&0.6784 &0.7050 &0.7083 &0.7148 &0.7156\\ 
&DMH&\bf 0.7012 &\bf 0.7057 &\bf 0.7398 &\bf 0.7424 &\bf 0.7501 \\ \hline
\multirow{6}{*}{Text-Image}& CMSSH &0.6613 &0.6510 & 0.6756 & 0.6643 & 0.6471\\ 
&CVH& 0.6495 & 0.6213 & 0.6179 & 0.6050 & 0.5948\\ 
&SCM& 0.6194 & 0.6302 & 0.6377 & 0.6377 & 0.6417\\ 
&SePH& 0.6745 &0.6824 & 0.6917 &0.7059 &0.7110\\ 
&MDBE& 0.7521 & 0.7793 & 0.7894 & 0.7903 & 0.7919\\ 
&DMH&\bf  0.7629 &\bf  0.7817 &\bf  0.7962 &\bf  0.8301 &\bf  0.8470\\ \hline

\end{tabular}
\begin{tabular}{|c|c|c	c	c	c	c|}
\hline
\multirow{ 2}{*}{Task} & \multirow{2}{*}{Method} & \multicolumn{5}{|c|}{\bf NUS-WIDE}\\ \cline{3-7} & & 16 bits & 32 bits & 64 bits & 96 bits & 128 bits \\ \hline
\multirow{6}{*}{Image-Text}& CMSSH & 0.4124 & 0.3533 & 0.3540 & 0.3578 & 0.3600\\ 
&CVH  & 0.4733 & 0.3505 & 0.2900 & 0.2812 & 0.2950\\ 
&SCM & 0.5245 & 0.5394 & 0.5332 & 0.5376 & 0.5400\\ 
&SePH  & 0.5573 &0.5481 & 0.5589 & 0.5572 & 0.5569\\ 
&MDBE & 0.6281 & 0.6409 & 0.6617 & 0.6598 & 0.6644\\ 
&DMH &\bf 0.6317 &\bf 0.6506 &\bf  0.6591 & \bf 0.6740 &\bf  0.6837\\ \hline
\multirow{6}{*}{Text-Image}& CMSSH & 0.4152 & 0.3515 & 0.3510 & 0.3555 & 0.3556\\ 
&CVH & 0.4794 & 0.4195 & 0.3901 & 0.3552 & 0.3501\\ 
&SCM& 0.5127 &0.5214 & 0.5255 & 0.5302 & 0.5380\\ 
&SePH& 0.7185 &0.7258 & 0.7390 & 0.7455 & 0.7491\\ 
&MDBE& 0.7623 & 0.7737 & 0.7953 & 0.7973 & 0.7987\\ 
&DMH&\bf  0.7653 &\bf  0.7827 &\bf  0.8150 &\bf  0.8192 &\bf  0.8246\\ \hline

\end{tabular}
\end{center}
\end{table*}
\section{Experimental Results}
\label{sec:experiment}
In this section, we evaluate the retrieval performance and computational efficiency of the proposed method. First, we introduce the data sets, evaluation metrics and comparison methods. Then, two types of experiments - \emph{Hamming ranking} and \emph{hash lookup} were conducted. Finally, we analyze the convergence and computational efficiency.
\subsection{Data sets}
\textbf{MIRFlickr}~\cite{MIRFlickr} contains 25,000 entries each of which consists of 1 image, several textual tags and labels. Following literature~\cite{SePH}, we only keep those textural tags appearing at least 20 times and remove entries which have no label. Hence, 20,015 entries are left. For each entry, the image is represented by a 512-dimensional GIST~\cite{GIST} descriptors and the text is represented by a 500-dimensional feature vector derived from PCA on index vectors of the textural tags. 5\% entries are randomly selected for testing and the remaining entries are used as training set. Ground-truth semantic neighbors for a test entry, i.e, a query, are defined as those sharing at least one label.\\
\indent \textbf{NUS-WIDE}~\cite{NUS-WIDE} is comprised of 269,648 images and over 5,000 textural tags collected from Flickr. Ground-truth of 81 concepts is provided for the entire data set. Following literatures~\cite{CMFH}\cite{SePH}\cite{SCM}, we select 10 most common concepts for labels and thus 186,577 entries are left. For each entry, the image is represented as a 500-dimensional bag-of-visual-words SIFT feature vector and text is represented as an index vector of the most frequent 1,000 tags. 1\% entries are randomly selected for testing and the remaining are used for training. Ground-truth semantic neighbors for a test entry are defined as those sharing at least one label.
\subsection{Evaluation Metrics}
\emph{Hamming ranking} and \emph{hash lookup} are two widely used experiments for evaluating retrieval performance. In Hamming ranking experiment, all data points in the training set are ranked depending on their Hamming distances to a given query. The average precision (AP) is defined as
\begin{equation}
AP=\frac{1}{N}\sum_{r=1}^R{P(r)\delta(r)}
\end{equation}
where $N$ is the number of relevant instances in the retrieved set, $P(r)$ is the precision of the top $r$ retrieved instances, and $\delta(r)=1$ if the $r$-th retrieved instance is a true neighbor of the query, and otherwise $\delta(r)=0$. Mean average precision (MAP) is the mean of APs of all the queries. For the ideal case that all retrieved instance are true neighbors of the queries, MAP is equal to 1, while MAP is equal to 0 for the worst case that all retrieved instance are not the true neighbors. Hence, the closer it is to 1, the better the performance.\\
\indent In \emph{hash lookup} experiment, the retrieved instances are those whose Hamming distances to a given query are not larger than a given radius, say 2 in our experiment. The performance are evaluated by F1-score which is defined as
\begin{equation}
F1=2\frac{precision\cdot recall}{precision+recall}
\end{equation}
The F1-scores are averaged for all queries. Similar to MAP, F1 also varies in $[0,1]$ and the closer it is to 1, the better the performance.
\subsection{Baselines}
The proposed method is compared with five state-of-the-art multimodal hashing methods CMSSH~\cite{CMSSH}, CVH~\cite{CVH}, MDBE~\cite{MDBE}, SCM~\cite{SCM} and SePH~\cite{SePH}.\\
\indent CMSSH and SePH requires too much computational cost. Following literatures~\cite{SePH}\cite{SCM}, 10,000 entries are randomly selected for training hashing functions and then we apply these functions to generate hashing codes. We use the codes provided by the authors except for MDBE. We re-implement MDBE and set parameters following the authors' suggestions. For our method, we use the following parameter settings, $k_s=0.003$, $k_e=0.0015$ and $K=400$. $\alpha_i$, $\beta_i$  and $\gamma_i$ are set as discussed in Subsection~\ref{subsec:id}.
\begin{figure*}[h]
\centering
\includegraphics[width=0.8\linewidth]{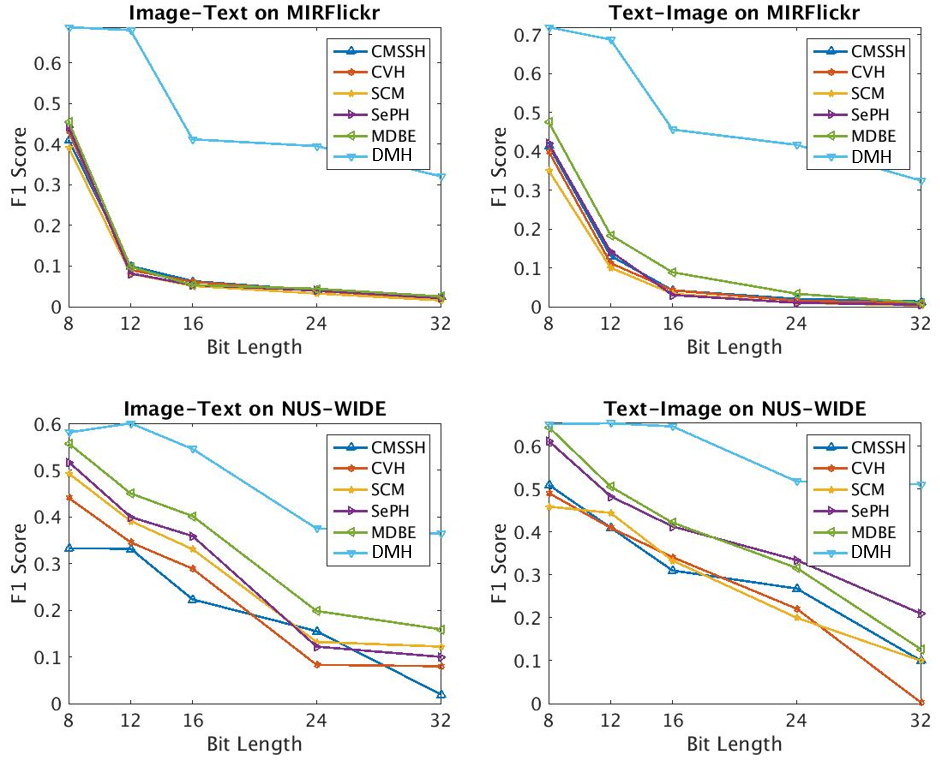}
\caption{F1-score on MIRFlickr and NUS-WIDE data sets}
\label{fig:f1}
\end{figure*}
\subsection{Results}
MAP results are shown in Table~\ref{tb:map}. In Table~\ref{tb:map}, ``Image-Text'' means using images to query texts, while ``Text-Image'' means using texts to query images. From Table~\ref{tb:map}, it can be observed that our method outperforms all compared methods. As the bit length increases, the performance of our method increases faster than baselines, which demonstrates the effectiveness of the proposed minimum correlation regularization. For example, in the ``Image-Text'' experiment on MIRFlickr, the performance improvement ranges from 3\% to 5\% as the bit length varies from 16 to 128, compared to the best baseline, i.e., MDBE. \\
\indent F1-score results are shown in Figure~\ref{fig:f1}. Similar to the MAP results, our method surpasses all baselines by a huge performance improvement, especially on MIRFlickr. On MIRFlickr, the performance improvement ranges from 30\% to 3,000\%, compared to the best baseline. On NUS-WIDE, it is 5\% to 200\%. A reasonable explanation is that our method can precisely preserve the inter-class structure and therefore the lookup performance is significantly improved. Because the ranking performance depends on the preservation of the structure of the whole data set regardless of inter-class or intra-class structure, it is not as significant as that of the lookup experiment. The size of MIRFlickr is only about 1/10 of NUS-WIDE, so the simple non-linearity introduced in our method works much better on MIRFlickr. To achieve comparable performance improvement on NUS-WIDE data set, more sophisticated non-linear models are expected.\\
\indent In both experiments, MDBE achieves the best performance among all the baselines. Actually, the main part of MDBE,
\begin{equation}\label{eq:mdbe}
\|\mathbf{LU}-\mathbf{XW_x}\|^2_F+\|\mathbf{LU}-\mathbf{YW_y}\|^2_F,
\end{equation}
is equivalent to Eq.~\eqref{eq:preModel} which is an intuitive multimodal extension of ITQ, where $L$ is the label matrix, $X$ is the image view matrix and $Y$ is the text view matrix. $W_x$, $W_y$ and $U$ are variables. If we treat the label matrix as another view of the data and introduce an auxiliary variable $B$, it is easy to figure out that Eq.~\eqref{eq:mdbe} and Eq.~\eqref{eq:preModel} are equivalent. By introducing non-linearity and minimum correlation regularization, our method performs much better than MDBE.\\
\begin{figure*}[h]
\centering
\includegraphics[width=0.9\linewidth]{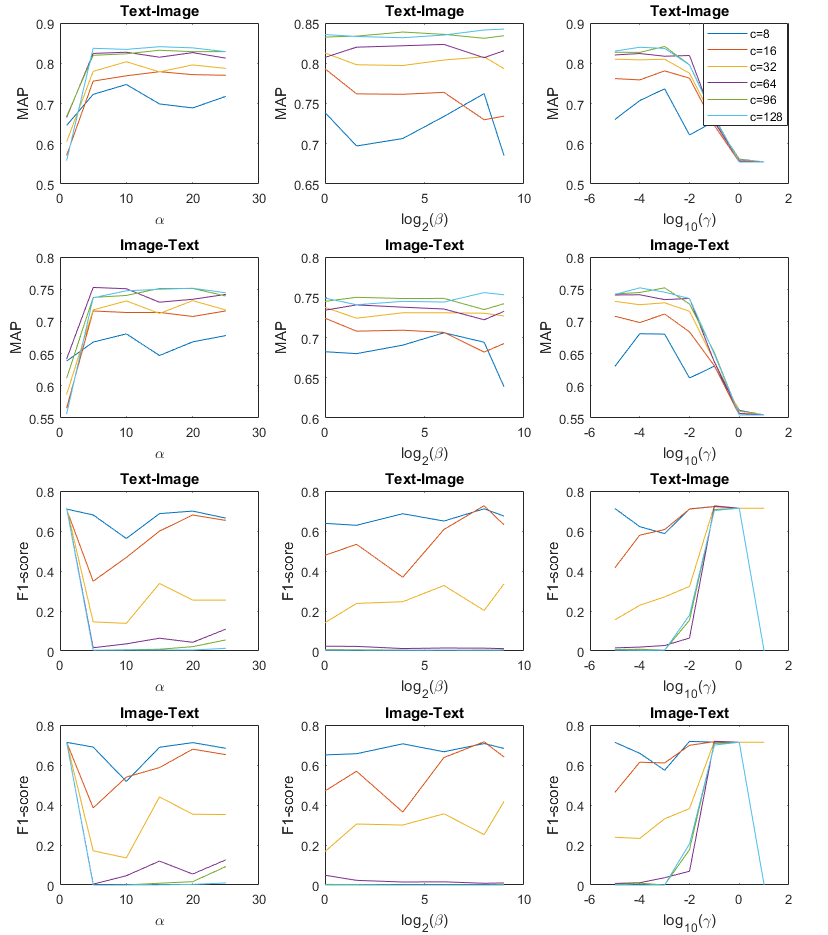}
\caption{MAP and F1-score of DMH on MIRFlickr data set.}
\label{fig:param}
\end{figure*}
\subsection{Parameter Settings}
In Fig.~\ref{fig:param}, we show the MAP and F1-score of DMH on MIRFlickr data set with various parameter settings. The default setting is $\alpha=10$, $\beta=255$ and $\gamma=0.001$. In each figure, only the tested parameter varies and the other two parameters keep their default values. \\
\indent In the left column of Fig.~\ref{fig:param}, $\alpha$ varies in $\{1,5,10,15,20,25\}$. It can be seen that the highest MAP is usually achieved by $\alpha=5$ or $\alpha=10$. The highest F1-score is got when $\alpha=1$. However, when $\alpha=1$, DMH performs badly in MAP. Hence, $\alpha=10$ is selected for our experiments to achieve a balanced performance on these two types of experiments. \\
\indent In the middle column of Fig.~\ref{fig:param}, $\beta$ varies in $2^{\{1,2,4,6,8,9\}}-1$. In the long-bit experiment ($c>16$), the performance is relatively robust to $\beta$. The highest F1-score is achieved when $\beta=255$. Hence, $\beta=255$ is used in our experiments.\\
\indent In the right column of Fig.~\ref{fig:param}, $\gamma$ varies in $10^{\{-5,-4,-3,-2,-1,0,1\}}$. It can be seen that DMH performs best in MAP when $\gamma=0.001$. When $\gamma>0.1$, F1-score rockets up, while MAP dumps. A possible explanation is that the regularization overly decorrelates a few columns of the code matrix and leaves other columns highly mutually correlated. The resulting code matrix will be similar to a short-bit code matrix. That is why MAP and F1-scores in all 6 experiments with different lengths of bits are rather close in this situation. Although the global optimum of MCR tends to generate column vectors similar to those illustrated in Fig.~\ref{fig:proposition}, the gradient descent algorithm cannot guarantee such solutions since MCR is not convex. Hence, $\gamma=0.001$ is used in our experiments.\\
\subsection{Convergence Study}
\begin{figure}[t]
\centering
\includegraphics[width=0.9\linewidth]{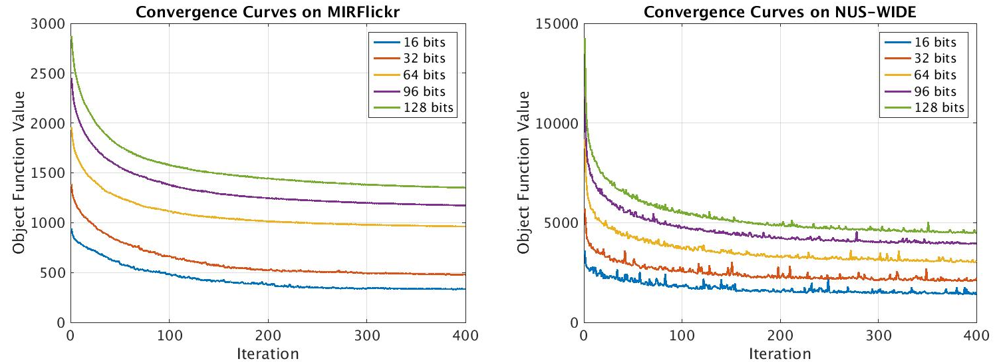}
\caption{Convergence curves on MIRFlickr and NUS-WIDE data sets}
\label{fig:conv}
\end{figure}
The objective function of our method is minimized by \textbf{Algorithm}~\ref{alg:2}. In \textbf{Algorithm}~\ref{alg:2}, we empirically amend the derivative of $E$ with respect to $\mathbf{W}^i$ for easy parameter tuning. The convergence property is experimentally studied in this subsection. Fig.~\ref{fig:conv} shows the convergence curves. It can be seen that the objective function value decreases fast in the first 100 iterations and then slides relatively slowly. The convergence curves of experiments on MIRFlickr is smooth, while those of experiments on NUS-WIDE jitters because of more sophisticated data structure and therefore more saddle points across which the algorithm jumps.
\subsection{Computation Efficiency}
Training and testing time on 32-bit are given in Table~\ref{tb:time}. The training time is the mean time of 10 runs. The testing time is the average time cost for one query. All experiments were performed on MATLAB R2015b installed on a GNU/Linux Server with 2.30 GHz 16-core CPU and 768 GB RAM. From Table~\ref{tb:time}, it can be seen that the training time of our method is moderate among all methods. Its testing time is close to that of MDBE, because the encoding procedure for a new query of these two methods are similar.
\begin{table}[h]
\caption{Training and Testing Time on MIRFlickr and NUS-WIDE data sets in seconds. The testing time is multiplied with $10^{-5}$}
\label{tb:time}
\begin{center}
\begin{tabular}{|c|c|c|c|c|}
\hline
 & \multicolumn{2}{|c|}{\textbf{MIRFlickr}} & \multicolumn{2}{|c|}{\textbf{NUS-WIDE}}\\ \hline
Method & Training & Testing & Training & Testing \\ \hline
CMSSH & 69.7 & 1.016 & 705.2 & 1.270  \\ \hline
CVH & 0.9 & 0.910 & 3.6  & 1.087 \\ \hline
SCM & 1.3 & 0.308 & 12.5 & 1.270 \\ \hline
SePH & 4711.2 & 4.244 & 5082.3 & 5.550 \\ \hline
MDBE & 25.0 & 0.431& 241.8 & 0.572\\ \hline
DMH & 29.8 & 0.432 &398.0 &0.572\\ \hline
\end{tabular}
\end{center}
\end{table}

\section{Conclusion}
\label{sec:conclusion}
This paper proposed an effective multimodal hashing method which is modeled as a quantization error problem and the minimum correlation regularization is devised to improve the retrieval performance on long codes. Experiments on MIRFlickr and NUS-WIDE data sets show that the proposed method surpasses the compared methods distinctively. Future works include testing more nonlinear embedding functions and refining optimization procedure for high computational efficiency.\\

\bibliographystyle{IEEEtranS}
\bibliography{Output4}

\end{document}